# Directional Characteristics of Thermal-Infrared Beaming from Atmosphereless Planetary Surfaces – A New Thermophysical Model

B. Rozitis[a] and S. F. Green[a]

[a]*Planetary and Space Sciences Research Institute, The Open University, Walton Hall, Milton Keynes, MK7 6AA, UK*

Please direct editorial correspondence and proofs to:

Benjamin Rozitis
Planetary and Space Sciences Research Institute
The Open University
Walton Hall
Milton Keynes
Buckinghamshire
MK7 6AA
UK

Phone: +44 (0) 1908 655808

Email: b.rozitis@open.ac.uk

Email address of co-author: s.f.green@open.ac.uk

**ABSTRACT**

We present a new rough-surface thermophysical model (Advanced Thermophysical Model or *ATPM*) that describes the observed directional thermal emission from any atmosphereless planetary surface. It explicitly incorporates partial shadowing, scattering of sunlight, selfheating and thermal-infrared beaming (re-radiation of absorbed sunlight back towards the Sun as a result of surface roughness). The model is verified by accurately reproducing ground-based directional thermal emission measurements of the lunar surface using surface properties that are consistent with the findings of the Apollo missions and roughness characterised by an RMS slope of ~32°. By considering the wide range of potential asteroid surface properties, the model implies a beaming effect that cannot be described by a simple parameter or function. It is highly dependent on the illumination and viewing angles as well as surface thermal properties and is predominantly caused by macroscopic rather than microscopic roughness. Roughness alters the effective Bond albedo and thermal inertia of the surface as well as moving the mean emission away from the surface normal. For accurate determination of surface properties from thermal-infrared observations of unresolved bodies or resolved surface elements, roughness must be explicitly modelled, preferably aided with thermal measurements at different emission angles and wavelengths.

# 1. INTRODUCTION

Planetary surfaces illuminated by the Sun are, on average, in equilibrium between the absorbed solar radiation and the thermal radiation emitted from the surfaces themselves (in the absence of significant internal heat sources). The instantaneous emitted thermal flux is dependent on the surface temperature distribution, which in turn is dependent on several factors associated with the planetary body. These include heliocentric distance, rotation rate, orientation of the spin vector, global shape, and a number of different surface properties including albedo, thermal inertia, and roughness. Thermal models combine shape and/or surface models with thermal physics to determine surface and/or sub-surface temperature distributions of atmosphereless planetary bodies. They are valuable tools for use within planetary science since they can be used to infer the above properties by comparing predicted thermal emission with remote sensing observations. They also permit investigations of the asteroid Yarkovsky and YORP effects, which are caused by the net force and torque resulting from asymmetric reflection and thermal re-radiation of sunlight from an asteroid's surface. The net force (Yarkovsky effect) causes the asteroid's orbital semi-major axis to change and the net torque (YORP effect) changes the asteroid's rotation period and the direction of its spin axis (Bottke et al. 2006). Prediction of these two effects, which are fundamental to the dynamical and physical evolution of small solar system bodies, is critically dependent on accurate thermal models.

The most useful properties for characterising an atmosphereless planetary surface include the thermal inertia and roughness. Since thermal inertia depends predominantly on regolith particle size and depth, degree of compaction, and exposure of solid rocks and boulders within the top few centimeters of the subsurface; it can be used to infer the presence or absence of loose material on the surface (Delbo' et al. 2007). It also dictates the strength of the asteroid Yarkovsky effect. Roughness can be defined as a measure of the irregularity of a surface at scales that are smaller than the global shape model resolution but larger than the thermal skin depth specified by the thermal inertia. Both properties significantly affect the observed planetary thermal emission.

Thermal inertia introduces a lag time between absorption and re-radiation of solar radiation. Increasing the thermal inertia decreases the day-side surface temperature distribution and increases it for the night-side. Roughness causes the surface to thermally emit in a non-lambertian way with a tendency to re-radiate the absorbed solar radiation back towards the Sun, an effect known as thermal-infrared beaming (Lagerros 1998). It is thought to be the result of two different processes: a rough surface will have elements orientated towards the Sun that become significantly hotter than a flat surface, and multiple scattering of radiation between rough surface elements increases the total amount of solar radiation absorbed by the surface.

There are two types of thermal model: simple and thermophysical. Simple thermal models using idealised (usually spherical) geometry and idealised assumptions of the level of thermal inertia and roughness, such as the Standard Thermal Model (STM) and Fast Rotating Model (FRM), have previously been used to determine asteroid diameters and albedos when simultaneous measurements of disc-integrated asteroid flux have been made in the visible and infrared (see Delbo' & Harris (2002) for a review). Although successful for determining diameters and albedos of main-belt asteroids, these models have obvious limitations when it comes to detailed interpretations from high quality spacecraft/observational data or for the prediction of accurate asteroid thermal infrared fluxes. This is especially true for near-Earth asteroids (NEAs) where they are known to exhibit much more irregular shapes than main-belt asteroids. The Near Earth Asteroid Thermal Model (NEATM by Harris (1998)) and the Night Emission Simulated Thermal Model (NESTM by Wolters & Green (2009)) attempt to account for thermal inertia and roughness for NEAs but still rely on idealised spherical geometry.

Thermophysical models use detailed shape and/or topography models with sophisticated thermal physics to make the model as realistic as possible. The very first thermophysical models were inspired by thermal infrared observations of the lunar surface conducted early in the 20th century that showed that the Moon emits thermal radiation in a non-Lambertian way (Pettit & Nicholson 1930). Thermophysical modelling of the lunar surface revealed that the lunar thermal-infrared beaming effect could be explained by considering the shadowing and mutual radiative heat exchange of various rough surfaces, in particular, a cratered surface in instantaneous equilibrium recreated the observed effect well (e.g. Smith 1967; Buhl, Welch & Rea 1968; Sexl et al. 1971; Winter & Krupp 1971).

With the success of the thermophysical models in their application to the lunar surface, their application to other planetary bodies was developed and their sophistication increased. The Spencer (1990) model for airless planets saw the first detailed treatment of thermal conduction within spherical section craters, of which Emery et al. (1998) produced a variant applicable to thermal-infrared observations of the planet Mercury. A two-surface-layer model including temperature-dependent thermal properties was produced by Vasavada, Paige & Wood (1999) for calculating the surface temperature distribution of specific regions on a planetary body where detailed topography models exist, and is currently in use for interpreting thermal-infrared observations of the lunar surface conducted by the Diviner instrument on the Lunar Reconnaissance Orbiter (Paige et al. 2010). Groussin et al. (2007) produced a smooth-surface model including the detailed 3D shape of the nucleus of Comet 9P/Tempel 1 to interpret spatially resolved thermal-infrared observations conducted by the Deep Impact spacecraft, and Davidsson, Gutiérrez & Rickman (2009) attempted to improve upon this by including surface roughness.

In asteroid science, the most commonly used thermophysical models are those produced by Johan Lagerros (Lagerros 1998), Marco Delbo' (Delbo' 2004), and Michael Müller (Müller 2007). All three models can represent an asteroid as an irregularly-shaped object split into a number of discrete surface elements (typically a few thousand), include shadowing and 1D heat conduction, and include mutual radiative-heat exchange within spherical-section craters. The way this is implemented differs slightly between the models. None of the models include temperature-dependent surface properties, multiple surface layers, or mutual radiative-heat exchange between interfacing global shape elements. Delbo's model can be seen as an update of the Spencer model to irregularly-shaped asteroids where spherical-section craters are split into a number of finite elements (typically ~40) and 1D heat conduction solved for each crater element. As 1D heat conduction has to be solved for each global shape and crater element the model has a relatively long run time. Also the low number of crater elements could cause inaccuracies in the emitted flux at high emission angles relative to the surface normal. Lagerros's model solves 1D heat conduction only for the global shape elements, and then determines the surface temperature distribution inside the craters analytically assuming no heat conduction. The thermal flux emitted from the crater is corrected by a ratio calculated by comparing the thermal flux from the global shape model element when it has non-zero heat conduction to zero heat conduction. The advantages with this model are that it is faster to run and that the thermal flux at high emission angles is potentially more accurate. However, it does come with one obvious disadvantage in that the rough surface thermal emission cannot be calculated on the night side of the asteroid, which of course can be done with Delbo's model. Müller's model is an update of Lagerros's model but does not solve the rough surface night emission problem.

Until only recently, the majority of thermal-infrared observations for these airless planetary bodies were disc-integrated, and so the majority of thermophysical models were developed only to investigate disc-integrated measurements. Relatively few were developed to investigate directionally- and spatially-resolved measurements that are expected to be gained from spacecraft. Other than the Apollo era lunar rough surface and the Comet 9P/Tempel 1 models the few other models developed include those by Colwell & Jakosky (2002) and Bandfield & Edwards (2008). Colwell & Jakosky considered spherical-section craters whilst Bandfield & Edwards considered a Gaussian distribution of surface slopes. These models were applied to spacecraft spatially-resolved thermal-infrared observations of specific regions on the lunar and martian surfaces respectively, and determined surface slopes that appeared consistent with the surface morphology seen in optical images of the same regions.

However, no model has been applied to investigate how thermal-infrared beaming varies with direction for spatially-resolved thermal emission as a function of the huge range of potential surface properties. A number of current and planned planetary space missions include thermal-infrared instruments to characterise the target's surface properties (e.g. Diviner on Lunar Reconnaissance Orbiter (Paige et al. 2010), VIRTIS on Rosetta (Coradini et al. 2007), and MERTIS on BepiColombo (Hiesinger, Helbert & MERTIS Co-I Team 2010)). Knowing how the surface thermal emission varies as a function of surface thermal properties and illumination and observation geometries will be useful in determining an appropriate spacecraft mapping strategy that maximises the amount of information that can be obtained about the surface.

We present here the implementation of a new model, called the Advanced Thermophysical Model (*ATPM*), to investigate the directionally-resolved thermal-infrared beaming effect. It is applicable to both spatially-resolved and disc-integrated measurements, and overcomes some of the limitations associated with previous thermophysical models. The model is initially verified by reproducing the directionally-resolved thermal-infrared observations of the lunar surface, and the inferred degree of roughness is then compared with that observed in images taken by the Apollo missions and by radar studies. The directional characteristics of thermal-infrared beaming are then studied for a generic asteroid surface.

In order to study thermal-infrared beaming in a directionally-resolved sense the illumination and observation geometry defined in Figure 1 is used. The illumination and observation angles, $\theta_{SUN}$ and $\theta_{OBS}$, are measured from the surface normal in a sense that conforms to conditions on Earth i.e. the Sun rises in the east and sets in the west. Morning angles are given negative values, and afternoon angles are given positive values.

## 2. THERMOPHYSICAL MODEL

### 2.1 Model Overview

Figure 2 displays a schematic giving a brief overview of the physics and geometry involved in the *ATPM*. The model accepts global shape models in the triangular facet formalism. It also accepts a topography model which it uses to represent the unresolved surface roughness in the global shape model for each facet. Any representation of the surface roughness can be used in the topography model but hemispherical craters are preferred since they are easy to parameterise. Both types of facet (shape and roughness) are considered large enough so that lateral heat conduction can be neglected and only 1D heat conduction perpendicular and into the surface can be considered. Therefore, for every shape and roughness facet a 1D heat conduction equation is solved throughout a specified number of planetary rotations with a surface boundary condition. The surface boundary condition includes direct and multiple-scattered solar radiation, shadowing, and re-absorbed thermal radiation from interfacing facets. The degree of surface roughness for the planetary body is specified by a roughness fraction, $f_R$, that dictates the fraction of the planetary body surface represented by the rough-surface shape model. The remaining fraction, $(1 - f_R)$, is represented by a smooth and flat surface. Finally, the observed thermal emission is determined by applying and summing the Planck function over every visible shape and roughness facet.

## 2.2 Thermal Physics

To determine the temperature $T$ for each facet the energy balance equation has to be solved. For each facet, conservation of energy leads to the surface boundary condition

$$(1 - A_B)((1 - S(t))\psi(t)F_{SUN} + F_{SCAT}) + (1 - A_{TH})F_{RAD} + k\left(\frac{dT}{dx}\right)_{x=0} - \varepsilon\sigma T^4_{x=0} = 0 \quad (1)$$

where $\varepsilon$ is the emissivity, $\sigma$ is the Stefan Boltzmann constant, $A_B$ is the Bond albedo, $S(t)$ indicates whether the facet is shadowed at time $t$, $k$ is the thermal conductivity, and $x$ is the depth below the planetary surface. $\Psi(t)$ is a function that returns the cosine of the Sun illumination angle at a time $t$, which depends on the facet and rotation pole orientations, and it changes periodically as the planetary body rotates. $F_{SUN}$ is the integrated solar flux at the distance of the object, which is given by $(1367 / r_H^2)$ W m$^{-2}$ where $r_H$ is the heliocentric distance of the planetary body in AU. Interfacing facets on an irregular planetary surface will receive an additional flux contribution from multiple-scattered sunlight and absorption of thermal emission from neighbouring facets. $F_{SCAT}$ and $F_{RAD}$ are then the total scattered and thermal-radiated fluxes incident on the facet respectively where $A_{TH}$ is the albedo of the surface at thermal-infrared wavelengths.

Heat conduction in the absence of an internal heat source can be described by the 1-D heat conduction (diffusion) equation

$$\frac{\partial T}{\partial t} = \frac{k}{\rho C}\frac{\partial^2 T}{\partial x^2} \quad (2)$$

where $k$, $C$, and $\rho$ are the thermal conductivity, specific heat capacity, and density of the surface material which for simplicity have been assumed to be constant with depth and temperature. Following the approach outlined by Wesselink (1948), if $\Psi(t)$ is considered to have a harmonic variation then it would produce a harmonic variation in surface temperature and also in internal temperature but with decreasing amplitude with depth such that it can be represented by

$$T(x,t) = a + b\exp\left(\frac{-2\pi x}{l_{2\pi}}\right)\cos\left(2\pi\left(\frac{t}{P_{ROT}} - \frac{x}{l_{2\pi}} + \xi\right)\right) \quad (3)$$

where $P_{ROT}$ is the rotation period of the planetary body, and $l_{2\pi}$ is the thermal skin depth at which the phase lag of the internal temperature variation is $2\pi$ and the amplitude of internal temperature variations has decreased by a factor $e^{-2\pi}$ and is given by

$$l_{2\pi} = \sqrt{\frac{4\pi P k}{\rho C}} \, . \quad (4)$$

This implies that equations (1) and (2) can be normalised using the new depth and time variables $z$ and $\tau$ given by

$$z = \frac{x}{l_{2\pi}} \qquad \tau = \frac{t}{P_{ROT}} \quad (5)$$

which transforms them into

$$(1 - A_B)((1 - S(\tau))\psi(\tau)F_{SUN} + F_{SCAT}) + (1 - A_{TH})F_{RAD} + \frac{\Gamma}{\sqrt{4\pi P_{ROT}}}\left(\frac{\partial T}{\partial z}\right)_{z=0} - \varepsilon\sigma T^4_{z=0} = 0 \quad (6)$$

$$\frac{\partial T}{\partial \tau} = \frac{1}{4\pi} \frac{\partial^2 T}{\partial z^2} \tag{7}$$

where $\Gamma$ is the surface thermal inertia and is given by

$$\Gamma = \sqrt{k\rho C} \ . \tag{8}$$

Since the amplitude of internal temperature variations decreases exponentially with depth it implies an internal boundary condition given by

$$\left(\frac{\partial T}{\partial z}\right)_{z \to \infty} \to 0 \ . \tag{9}$$

A finite difference numerical technique is used to solve the problem defined by equations 6, 7, and 9. If $T_{i,j}$ is the temperature at depth $z = i.\delta z$ and rotation phase $\tau = j.\delta \tau$ (for $i = 1$ to $n$ depth steps and $j = 1$ to $m$ time steps) then equation 7 becomes the following after rearranging for $T_{i,j+1}$

$$T_{i,j+1} = T_{i,j} + \frac{1}{4\pi} \frac{\delta \tau}{(\delta z)^2} \left[T_{i+1,j} - 2T_{i,j} + T_{i-1,j}\right] . \tag{10}$$

However, this does not allow determination of $T_{0,j+1}$ or $T_{n,j+1}$ which require exploiting the boundary conditions 6 and 9. In terms of difference equations the internal boundary condition becomes

$$T_{n,j+1} = T_{n-1,j+1} \ . \tag{11}$$

To transform the surface boundary condition into difference equation terms the following substitution is made

$$\left(\frac{\partial T}{\partial z}\right)_{z=0} = \frac{1}{\delta z}\left[T_{1,j+1} - T_{0,j+1}\right] . \tag{12}$$

The surface boundary condition now contains a derivative with respect to $z$ and the surface temperature itself. This can be solved using an iterative technique such as Newton-Raphson i.e. if $T_R$ is an approximate solution of $f(T_R) = 0$ then a closer approximation is given by

$$T_{R+1} = T_R - \frac{f(T_R)}{f'(T_R)} \ . \tag{13}$$

**2.3 Shadowing, Multiple Sunlight Scattering, and Re-absorption of Thermal Radiation**

Two types of shadowing occur on a planetary surface: horizon shadows where the Sun dips below the local horizon, and projected shadows where a facet gets in the way of another facet's line of sight to the Sun. A facet is considered to be horizon shadowed if its illumination angle, i.e. the angle between the facet's normal and line of sight to the Sun, is greater than or equal to 90º. Projected shadows are more difficult to determine and are calculated by using a ray-triangle intersection method to determine whether a facet is shadowed by another. A triangular facet is by its definition part of a much larger plane that is defined by its three vertices but is limited by its three edges. The ray triangle intersection is performed in two steps: firstly the direct sunlight ray on a test facet is intersected with a shadow-casting facet's plane, and secondly it is checked whether the intersection is made within the boundaries of the shadow-casting facet. If so, then the test facet is considered to be shadowed. To check for shadows formed across the entire surface each facet has to be tested for projected shadows with every other facet. This is a computational $N^2$ problem but it only needs to be performed once in each situation since the results can be saved to and reused from a lookup table.

Generally, for an illuminated facet $S(\tau) = 0$ and for a shadowed facet $S(\tau) = 1$. However, depending on the resolution of the shape models used the shadow tests described above can become inaccurate in certain situations. For example, the shadow cast by one facet could fall on half the area of another facet but due to the binary nature of the shadow tests described above the facet which is half shadowed will either be determined to be fully shadowed or not shadowed at all. To ensure shadowing accuracy, the highest resolution shape models should be used to minimise this effect. However, the topography models used to represent unresolved surface roughness must be of the lowest possible resolution to minimise the model run time. A compromise can be achieved by measuring the fraction of the area that is shadowed for each facet allowing the direct solar illumination imposed on each facet to be reduced accordingly. To determine the area fraction under shadow for a particular facet it can be divided up into a number of equal-area subfacets, $MM$, (typically 100) and the shadow tests are performed on each subfacet assuming shadows are cast by the full-size facets. A partial shadow fraction for each full-size facet can then be determined by summing the results of the subfacet shadow tests and dividing by the number of subfacets in each full-size facet

$$S(\tau) = \frac{1}{MM} \sum_{k=1}^{MM} s_k(\tau) \ . \tag{14}$$

Interfacing facets on an irregular planetary surface will receive additional flux contributions from multiple-scattered sunlight and reabsorbed thermal radiation. This exchange of heat between facets presents a radiative heat transfer problem, which is solved by using viewfactors. The viewfactor from facet $i$ to facet $j$, $f_{i,j}$, is defined as the fraction of the radiative energy leaving facet $i$ which is received by facet $j$ assuming Lambertian emission (Lagerros 1998). It is

$$f_{i,j} = v_{i,j} \frac{\cos\theta_i \cos\theta_j}{\pi d_{i,j}^2} a_j \tag{15}$$

where $v_{i,j}$ indicates whether there is line-of-sight visibility between the two facets, $\theta_i$ is facet $i$'s emission angle, $\theta_j$ is facets $j$'s incidence angle, $d_{i,j}$ is the distance separating facet $i$ and $j$, and finally $a_j$ is the surface area of facet $j$. The inter-facet visibility is again determined by the shadowing tests described above, and the results can be saved to a lookup table.

The viewfactor given by equation 15 is an approximation since it applies to large separation distances relative to the facet area. It can become very inaccurate when the relative separation distances are very small and can even produce a viewfactor greater than 1 which will obviously not conserve energy. A simple method to calculate the viewfactor between any two facets that fail the approximation criteria is to split them up into a number of equal-area subfacets (in the same manner as for partial shadowing above), $MM$, and determine the viewfactors associated with each subfacet combination. The effective overall viewfactor in this case is given by

$$f_{i,j} = \frac{1}{a_i} \sum_{v=1}^{MM} \left( a_{iv} \sum_{u=1}^{MM} f_{iv,ju} \right) \tag{16}$$

where $a_{iv}$ is the area of subfacet $iv$ which is part of facet $i$, and $f_{iv,ju}$ is the viewfactor from subfacet $iv$ to subfacet $ju$ as calculated by equation 15.

If only single scattering of sunlight is considered then the scattered sunlight flux contribution for facet $i$, $F_{SCAT}(\tau)$, is

$$F_{SCAT}(\tau) = A_B \cdot F_{SUN} \sum_{j \neq i} f_{i,j} \left(1 - S_j(\tau)\right) \psi_j(\tau) \tag{17}$$

where $S_j(\tau)$ indicates whether facet $j$ is shadowed at time $\tau$, and $\Psi_j(\tau)$ gives the cosine of the Sun illumination angle for facet $j$ at time $\tau$. Single scattering is a good approximation for low Bond albedos, although for high Bond albedos where multiple scattering occurs more easily it is less so. The scattered flux leaving facet $i$, $G_i(\tau)$, can be written as

$$G_i(\tau) = A_B \cdot \left( F_{SUN}(1-S_i(\tau))\psi_i(\tau) + \sum_{j \neq i} f_{i,j} G_j(\tau) \right) \quad (18)$$

which can be efficiently solved using the Gauss-Seidel iteration

$$G_i^{k+1}(\tau) = A_B \cdot \left( F_{SUN}(1-S_i(\tau))\psi_i(\tau) + \sum_{j>i} f_{i,j} G_j^k(\tau) + \sum_{j<i} f_{i,j} G_j^{k+1}(\tau) \right) \quad (19).$$

After a suitable number of iterations the multiple scattered flux incident on a facet is then

$$F_{SCAT}(\tau) = \frac{G(\tau)}{A_B} \quad (20).$$

For quick convergence to a solution the Gauss-Seidel iteration requires a suitable starting point close to the solution. In this case the single scattered derived fluxes can be used.

Every facet will receive thermal flux from visible interfacing facets with non-zero temperatures. The total incident thermal flux contribution for facet $i$, $F_{RAD}(\tau)$, is then a summation over all visible facets

$$F_{RAD}(\tau) = \varepsilon \sigma (1 - A_{TH}) \sum_{j \neq i} f_{i,j} T_j^4(\tau) \quad (21)$$

where $T_j(\tau)$ is the surface temperature of facet $j$ at time $\tau$. Single scattering is only considered since at thermal-infrared wavelengths planetary surfaces absorb most of the incoming radiation, i.e. $A_{TH} \sim 0$, and is a good approximation.

## 2.4 Thermal Emission Spectra

When the temperature $T_i(\tau)$ at time $\tau$ for a facet is known, the intensity of radiation it emits $I_{\lambda,i}(\tau)$ at a desired wavelength $\lambda$ is given by the Planck function

$$I_{\lambda,i}(\tau) = \frac{2\pi h c^2}{\lambda^5} \frac{1}{\exp\left(\frac{hc}{\lambda k T_i(\tau)}\right) - 1} \quad (22)$$

where $h$ is the Planck constant, $c$ is the speed of light, and $k$ is Boltzmann's constant. The spectral flux seen by an observer $F_{\lambda,i}(\tau)$ from facet $i$ assuming Lambertian emission is then

$$F_{\lambda,i}(\tau) = I_{\lambda,i}(\tau) \frac{a_i}{\pi d_i^2} \cos \theta_i \quad (23)$$

where $a_i$ is the area of the facet, $d_i$ is the distance to the observer, and $\theta_i$ is the observation angle measured away from the surface normal. However, the flux seen by an observer is a sum of fluxes from all shape and roughness facets visible within their field of view, and is given by

$$F_\lambda(\tau) = \sum_{i=1}^N v_i(\tau) \left( (1-f_R) F_{\lambda,i}(\tau) + ACF \cdot f_R \sum_{j=1}^M v_{ij}(\tau) F_{\lambda,ij}(\tau) \right) \quad (24)$$

where $v_i(\tau)$ and $v_{ij}(\tau)$ indicates whether the shape or roughness facet is visible respectively, and $f_R$ denotes the fraction of the surface represented by the rough-surface shape model (for $i = 1$ to $N$ shape facets and $j = 1$ to $M$ roughness facets). The facet visibility can be determined using the exact same method for shadowing (including the method for partial shadowing). The *ACF* term is an area conversion factor since the roughness topography model may not necessarily have the same spatial units as the global shape model (see Appendix A).

**2.5 Model Implementation**

In order to determine the illumination and observation geometries for accurate calculation of the incident and thermal-emission fluxes, a set of five related coordinate systems were specified. These are the heliocentric ecliptic, planetcentric ecliptic, planetcentric equatorial, and co-rotating planetcentric equatorial coordinate systems for specifying the global shape and orientation of a planetary body in space, and the surface-roughness coordinate system for specifying the unresolved surface topography. These coordinate systems and their relations are described in more detail in Appendix A. In each coordinate system the geometry of each triangular facet can be determined using its three vertices. In particular, the facet normal, $\boldsymbol{n}$, can be found by

$$\boldsymbol{n} = (\boldsymbol{p_1} - \boldsymbol{p_0}) \times (\boldsymbol{p_2} - \boldsymbol{p_0}) \qquad (25)$$

where $\boldsymbol{p_0}$, $\boldsymbol{p_1}$, and $\boldsymbol{p_2}$ are position vectors of the facet's three vertices which have been defined in an anti-clockwise sense so that the facet's normal points outwards from the closed surface. The area of the facet, $a$, can be found by

$$a = \frac{|\boldsymbol{n}|}{2} \qquad (26)$$

and the facet midpoint, $\boldsymbol{p_{mid}}$, can be found by

$$\boldsymbol{p_{mid}} = \frac{\boldsymbol{p_0} + \boldsymbol{p_1} + \boldsymbol{p_2}}{3} \quad . \qquad (27)$$

Various angles of interest $\theta$, such as the illumination and observation angles, can be found by utilising the dot product rule with the surface normal and the vector of interest $\boldsymbol{I}$

$$\boldsymbol{I} \cdot \boldsymbol{n} = |\boldsymbol{I}||\boldsymbol{n}|\cos\theta \quad . \qquad (28)$$

Appropriate values and settings should be assigned to the various parameters outlined in the previous sections for correct functioning of the model, the first being the number of time and depth steps the finite-difference technique should use and to what depth the 1D heat conduction equation should be solved. The thermal skin depth given by equation 4 gives the depth at which diurnal temperature variations have decreased by a factor of $e^{-2\pi}$ or $\sim 10^{-3}$, which becomes $\sim 10^{-6}$ for two thermal skin depths. For comparison purposes, previous thermophysical models tend to refer to the thermal skin depth as the depth at which diurnal temperature variations have decreased by a factor $e^{-1}$, $l_1$, given by

$$l_1 = \sqrt{\frac{kP_{ROT}}{2\pi\rho C}} \quad . \qquad (29)$$

The number of time and depth steps chosen should be high enough such that diurnal and depth temperature variations are easily resolved. However, for stability the finite-difference numerical technique suffers from the limitation

$$\frac{1}{4\pi}\frac{\delta\tau}{(\delta z)^2} < 0.5 \qquad (30)$$

which places constraints on the values chosen. The model uses as a default 400 time steps and 60 depth steps going down to a maximum depth of 2 thermal skin depths, which gives sufficient resolution, maintains accuracy at maximum depth, and easily avoids the limitation.

In order for the model to execute, it requires initialisation and it also needs to know when to stop. For rapid convergence to a solution, the initial temperature distribution must be chosen so that $T$ at large depths is close to the final solution, since it will take a long time for the surface changes to propagate to the centre. As a simple starting point, zero heat conduction is assumed and reabsorbed thermal radiation neglected so that the mean surface temperature, $\langle T_{z=0}\rangle_1$, across a whole rotation period can be calculated by

$$\langle T_{z=0}\rangle_1 = \left(\frac{(1-A_B)}{\varepsilon\sigma}\right)^{1/4} \frac{\int_{\tau=0}^{1}\left((1-S(\tau))\psi(\tau)F_{SUN} + F_{SCAT}\right)^{1/4}d\tau}{\int_{\tau=0}^{1}d\tau} \qquad (31)$$

where $F_{SCAT}$ has been calculated by the Gauss-Seidel iteration given above to an accuracy goal of 0.001 W m$^{-2}$. However, if there are interfacing facets then a better initial temperature distribution, $\langle T_{z=0}\rangle_2$, can be obtained by including reabsorbed thermal radiation

$$\langle T_{z=0}\rangle_2 = \left(\frac{1}{\varepsilon\sigma}\right)^{1/4} \frac{\int_{\tau=0}^{1}\left((1-A_B)((1-S(\tau))\psi(\tau)F_{SUN} + F_{SCAT}) + (1-A_{TH})F_{RAD}\langle T_{z=0}\rangle_1\right)^{1/4}d\tau}{\int_{\tau=0}^{1}d\tau} \qquad (32)$$

where the $F_{RAD}\langle T_{z=0}\rangle_1$ component is based on the mean surface temperature obtained by the first initialisation step. The initial temperature at all depths is then set equal to the mean surface temperature.

Knowing when to stop can be a bit more tricky as the model needs to execute quickly but must also maintain accuracy. As the model comes closer to a solution after each revolution the difference in surface temperature between consecutive revolutions decreases. Therefore, a simple and easy way to know when to stop the model is when the surface temperature difference between consecutive revolutions becomes less than a certain accuracy value $T_{ACC}$

$$T(\tau) - T(\tau-1) < T_{ACC} \qquad (33)$$

where $T(\tau)$ and $T(\tau\text{-}1)$ are the surface temperature distributions for the model's current and previous revolutions respectively. The result of the Newton-Raphson technique for solving the surface boundary condition must have sufficient accuracy so that the above convergence criteria can be applied. To ensure this, the convergence requirement for the Newton-Raphson iteration is when the temperature difference between consecutive iterations becomes less than one tenth of $T_{ACC}$

$$T_{r+1} - T_r < \frac{T_{ACC}}{10} \qquad (34).$$

However, the rate at which the model convergences is highly dependent on the thermal inertia value. Models with low thermal inertia converge quickly and the temperature differences between revolutions are relatively large, whereas those with high thermal inertia converge slowly and the temperature differences between revolutions are relatively small. A more accurate way of knowing when the model has converged is by checking the model's energy conservation fraction, $E_{CONS}$, given by

$$E_{CONS} = \frac{E_{OUTPUT}}{E_{INPUT}} \qquad (35)$$

where $E_{OUTPUT}$ is the total thermal radiation energy output of the planetary surface less the total amount of reabsorbed emitted thermal radiation, and $E_{INPUT}$ is the total sunlight absorbed by the surface taking into account multiple scattering of sunlight, both summed over one planetary rotation. A typical energy conservation goal for the model would be $0.97 < E_{CONS} < 1.0$, which is achievable for low thermal inertias ($\Gamma < 750$ J m$^{-2}$ K$^{-1}$ s$^{-1/2}$) with a $T_{ACC}$ of 0.05 K. However, for high thermal inertias ($\Gamma > 750$ J m$^{-2}$ K$^{-1}$ s$^{-1/2}$) with the same $T_{ACC}$ then $E_{CONS}$ becomes ~0.9 to 0.95. To ensure the same degree of energy conservation for high thermal inertias then a $T_{ACC}$ of 0.025 K is required, which also requires more model iterations to converge and therefore a longer model run time. To minimise the run time it is possible for the model to iterate only on shape and roughness facets that hadn't converged in previous iterations.

The model code was written in Microsoft Visual Studio 2008 Professional Edition in C++ to take advantage of object orientated programming, and 64bit and parallel computing. The model comprises several programs that each have a specific task in the thermal modelling process and output an appropriate lookup table that can be used by the next program. It is split up into the following stages: shape model generation, shadow map generation, selfheating map generation, thermal modelling, visibility map generation, observation modelling, and result rendering.

## 2.6 Surface Roughness Representations

Depending on the spatial scale at which you observe a planetary surface you may see craters and depressions, hills and mountains, rocks and boulders, pebbles and stones, powders, valleys, smooth flat surfaces, or more likely a mixture of all of them. Jakosky, Finiol & Henderson (1990) studied the thermal-infrared beaming effect caused by microscopic roughness, i.e. roughness at spatial scales smaller than the thermal skin depth, via experimental and theoretical directional emissivity studies of smooth playa and sand surfaces. They found that the thermal emission profile behaved more and more like a Lambert emitter with increasing microscopic roughness and found that only very smooth surfaces caused the thermal emission to be directed more towards the surface normal. Since all planetary surfaces have a microscopic rough surface, it follows that predominantly macroscopic roughness (occurring at spatial scales larger than the thermal skin depth) causes thermal-infrared beaming. This implies that microscopic beaming can be neglected from thermophysical models.

The work presented here utilises spherical-section craters of various opening angles, Gaussian random-height surfaces, and a flat surface in order to induce thermal-infrared beaming and to compare their results. Various resolutions of a 90º crater are also utilised to determine the effectiveness of the partial shadowing and visibility techniques introduced in the previous sections. The highest resolution crater model was designed to minimise shadowing errors at high illumination and observation angles by having an increased shape resolution around its rim. Therefore, this model does not require the partial shadowing and visibility tests and provides a good benchmark for the lower resolution models that would be using them. Figure 3 displays wireframe renderings of these rough surfaces and Table 1 lists their surface properties.

The roughness of a surface is measured in terms of the root-mean-square (RMS) slope. It is defined by weighting the square of the angular slope of each facet, including any flat facets, by its area projected on a local horizontal surface (Spencer 1990). The maximum RMS slope, $\theta_{MAX\_RMS}$, of the rough surface topography models can be calculated by

$$\theta_{MAX\_RMS} = \sqrt{\frac{\sum_{j=1}^{M} \theta_{S,j}^2 a_j \cos\theta_{S,j}}{\sum_{j=1}^{M} a_j \cos\theta_{S,j}}} \qquad (36)$$

where $\theta_j$ is the angle between roughness facet $j$'s normal and the normal of the local horizontal surface (for $j = 1$ to $M$ roughness facets). Since the roughness fraction, $f_R$, specifies the fraction of the planetary body surface represented by the rough-surface shape model and the remaining fraction represents a smooth flat surface, the overall roughness of the planetary surface, $\theta_{RMS}$, can be calculated by

$$\theta_{RMS} = \sqrt{f_R} \cdot \theta_{MAX\_RMS}. \tag{37}$$

Furthermore, the amount of selfheating that occurs within a rough surface can be measured in terms of the mean total viewfactor, $t_{view}$, which gives an indication of the degree of obscuration of any given facet's sky by other parts of the rough surface. It can be calculated by

$$t_{view} = \frac{1}{M} \sum_{i=1}^{M} \sum_{j \neq i} f_{i,j}. \tag{38}$$

## 3. LUNAR VERIFICATION

### 3.1 The Data

Saari & Shorthill (1972) obtained 23 scans of the sunlit portion of the lunar surface throughout a lunation. Observations were simultaneously conducted at wavelengths 0.45 and 10-12 μm to a spatial resolution of about 1% of the lunar radius (~18 km). They used the data to produce an isothermal and isophotic atlas of the Moon to study its albedo and surface brightness temperature statistics. Saari, Shorthill & Winter (1972) then used the atlas for directional emission studies by extracting surface brightness temperatures at a number of phase angles along the lunar equator. They present the data in two different graphical forms that are useful for the verification of the *ATPM*. The first set of graphs (Figures 1 to 3 of Saari, Shorthill & Winter (1972)) present surface brightness temperatures measured at fixed observation angles ($\theta_{OBS}$ = 0°, ±30°, and ±53°) as a function of Sun angle ($\theta_{SUN}$ = -90° to 90°). The second set of graphs (Figures 5 to 12 of Saari, Shorthill & Winter (1972)) present directional factor, $D$, at fixed Sun angles ($\theta_{SUN}$ = ±10°, ±20°, ±30°, ±40°, ±50°, ±60°, ±70°, and ±80°) as a function of observation angle ($\theta_{OBS}$ = -90° to 90°). The directional factor is defined as the ratio of the observed surface brightness temperature, $T_B$, of a specific region located $\theta_{SSP}$ degrees from the subsolar point to the expected temperature of a Lambertian surface, $T_L$,

$$D = \frac{T_B}{T_L} \tag{39}$$

where $T_L$ is calculated by

$$T_L = \left( \frac{F_{SUN}(1-A_B)}{\varepsilon \sigma} \right)^{1/4} \cos^{1/4} \theta_{SSP}. \tag{40}$$

Unfortunately the data points on these graphs have no associated error bars and so the measurement uncertainties are unknown.

### 3.2 Model Testing

The advantage with testing the *ATPM* model with thermal-infrared observations of the lunar surface is that all of the input parameters required for the model have already been measured by *in-situ* studies, particularly during the Apollo program. Therefore, inputting these measured parameters should cause the model to exactly reproduce the thermal-infrared observations at the appropriate level of surface roughness if it provides a good representation of a rough surface. The input model parameters were chosen as described below and are summarised in Table 2.

The thermal conductivity of the lunar surface was studied *in-situ* by heat flow experiments left by the Apollo 15 and 17 astronauts (Keihm et al. 1973; Keihm & Langseth 1973). These experiments found the surface to consist of multiple layers: a highly insulating top layer ~2 cm thick with a thermal conductivity ~$1 \times 10^{-3}$ W m$^{-1}$ K$^{-1}$, another layer below depths of ~10 cm with an increased thermal conductivity of ~$1 \times 10^{-2}$ W m$^{-1}$ K$^{-1}$, and a gradual transition between the two at the intermediate depths. The thermal conductivity was also found to be highly temperature-dependent suggesting that 70% of the heat exchange between regolith grains is radiative rather than conductive. The thermal properties of returned soil samples studied in the laboratory were also found to be highly temperature-dependent with measured heat conductivities similar to those measured *in-situ* (Linsky 1973 and references therein). The specific heat capacity of a returned Apollo 11 rock and soil sample was measured to be ~875 J kg$^{-1}$ K$^{-1}$ and likewise was found to be temperature-dependent (Robie, Hemingway & Wilson 1970). Also, the soil density in the ambient conditions of the upper 10 cm of the lunar surface was determined to range from 1300 to 1640 kg m$^{-3}$ (Linsky 1973). All of these studies indicate that the thermal inertia of the lunar surface is very low but assigning an exact value is complicated by multiple layers and temperature-dependent thermal properties. However, since the directional thermal emission studies were conducted on the sunlit side of the Moon, the lunar surface can be approximated by a single layer and a fixed thermal properties model. This is a valid approximation since heating by solar radiation dominates over sub-surface heat conduction on the Moon's sunlit side. Keihm et al. (1973) showed that the heat flow from the lower depths doesn't contribute significantly to the surface thermal emission until 15° of rotation phase after sunset, and Urquhart & Jakosky (1997) showed that the temperature-dependency of the thermal properties only became important on the night side of the Moon. Therefore, a thermal inertia, consistent with the measured thermal properties listed above, of 50 J m$^{-2}$ K$^{-1}$ s$^{-1/2}$ was assumed for the lunar surface in the model.

A model Bond albedo of 0.1 was assumed as Saari & Shorthill (1972) found it to vary across the lunar disc between the values of 0.065 and 0.276 with a mean value of 0.122. The model emissivity was assumed to be 0.9 as the multiple measurement attempts via different techniques listed in Linsky (1973) found it to vary between the values of 0.85 and 0.93 with a mean value of 0.89. A thermal albedo of 0.1 (i.e. 1 - $\varepsilon$) was also assumed.

Even though the Moon's orbit about the Earth is inclined to the ecliptic plane by ~5° its rotation axis is inclined to the same plane by only ~1°. This means that the Moon can be approximated very well for accurate calculation of equatorial surface temperatures by assuming a rotation pole orientation perpendicular to the ecliptic plane. Finally, the Moon's synodic period is taken as the time it takes for a surface element on the equator to be rotated back into the same solar illumination geometry.

A thermal model was run for each of the rough surfaces presented in Figure 3 and a flat surface, using the parameters listed in Table 2. The directionally resolved flux averaged over the wavelengths 10 to 12 μm at a roughness fraction $f_R$ was calculated using the methods presented in section 2 assuming the observer was situated far enough away from the rough surface that it could be considered a point source. The corresponding surface brightness temperature at 11 μm (the central wavelength) as a function of roughness fraction, and observation and sun angles, $T_{B,MOD,11\mu m}(f_R, \theta_{OBS}, \theta_{SUN})$, was calculated from the model observed flux intensities, $I_{MOD,11\mu m}(f_R, \theta_{OBS}, \theta_{SUN})$, by inverting equation 22 and the corresponding direction factors, $D_{MOD,11\mu m}(f_R, \theta_{OBS}, \theta_{SUN})$, were calculated using equations 39 and 40. The best fitting roughness fraction was found by minimising the least squares difference between the model results and the observations ($T_{B,OBS,10-12\mu m}(\theta_{OBS}, \theta_{SUN})$ and $D_{OBS,10-12\mu m}(\theta_{OBS}, \theta_{SUN})$) normalised by the solution for a flat smooth surface:

$$\chi_B^2(f_R) = \frac{\sum \left( T_{B,MOD,11\mu m}(f_R, \theta_{OBS}, \theta_{SUN}) - T_{B,OBS,10-12\mu m}(\theta_{OBS}, \theta_{SUN}) \right)^2}{\sum \left( T_{B,MOD,11\mu m}(f_R = 0, \theta_{OBS}, \theta_{SUN}) - T_{B,OBS,10-12\mu m}(\theta_{OBS}, \theta_{SUN}) \right)^2} \qquad (41)$$

$$\chi_D^2(f_R) = \frac{\sum \left(D_{MOD,11\mu m}(f_R, \theta_{OBS}, \theta_{SUN}) - D_{OBS,10-12\mu m}(\theta_{OBS}, \theta_{SUN})\right)^2}{\sum \left(D_{MOD,11\mu m}(f_R = 0, \theta_{OBS}, \theta_{SUN}) - D_{OBS,10-12\mu m}(\theta_{OBS}, \theta_{SUN})\right)^2} \quad (42)$$

The roughness fractions at which these $\chi^2$ values were minimised indicate lunar surface roughness and give a corresponding RMS slope.

### 3.3 Lunar Model Results

Figures 4 and 5 display the model fits to the data using the medium-resolution 90° crater and indicate that a very good fit can be obtained. Table 3 summarises the minimum $\chi^2$ values and the corresponding RMS slope for each roughness representation for the two sets of observations. Each RMS slope angle has associated uncertainty limits which indicate the RMS slope angles where the $\chi^2$ value is 10% greater than its minimum. Other than a completely smooth and flat surface the worst-fitting rough surface is the 30° crater, even at 100% coverage. It is simply not rough enough and it can only indicate that a roughness greater than 20.9° of RMS slope is required. The next worst-fitting rough surface is the low-resolution Gaussian random height surface, presumably due to its very low number of shape facets (i.e. 200). In the middle of the $\chi^2$ value range are the 45° crater and the high-resolution Gaussian random height surface, although their corresponding RMS slopes differ by ~11°.

    The rough surfaces that have the lowest $\chi^2$ values include the 60° crater and the 90° craters of different resolutions with the 90° craters producing slightly lower values than the 60° crater. In this case, the corresponding RMS slopes differ by 2° to 5° but are overlapped by their uncertainties. The different resolutions of the 90° crater produce almost identical $\chi^2$ values and corresponding RMS slopes, which verifies that the partial shadowing and visibility techniques work well.

    A small consistent discrepancy between the model and data can be seen on the afternoon side near sunset and at large negative observation angles (i.e. $\theta_{SUN} > 50°$ and $\theta_{OBS} < -30°$). It could be caused by a systematic error in the measurements and their corrections, especially as these sets of measurements would have had low signal to noise. For example, Saari & Shorthill (1972) performed albedo corrections to the observed thermal flux from each region on the lunar surface using the local and lunar average albedos to allow comparison of thermal fluxes from different regions. Each observation angle corresponds to a specific location along the lunar equator because of the Moon's tidally locked rotation. These data points are located in a region with an albedo that is higher than the lunar average. If the local albedo used in the corrections was slightly inaccurate, it could lead to the consistent discrepancy seen between the model and data. Alternatively, it could be caused by the assumption of uniform surface thermal properties used in the model. For example, if the thermal inertia of these regions was lower than the lunar average then it would cause the model to over-predict the directional factors at these regions. However, since the data points have no associated error bars it is impossible to assess the level of discrepancy and determine its cause.

    Averaging the RMS slope results from the different roughness representations give the derived RMS slopes as 31.5 ± 1.5 and 33.0 ± 1.1 degrees for the two sets of observations. These values are consistent with lunar RMS slopes derived by previous thermal models (see Table 4). However, these previous thermal models only performed a fit to one sub-set of the Saari, Shorthill & Winter (1972) data whilst the *ATPM* presented here is fitted to every sub-set simultaneously.

### 3.4 Geological Interpretation of Derived Lunar Roughness

Other than Spencer (1990) none of the previous thermal models compare their derived RMS slopes with other measurements of surface roughness made by alternative techniques. Primarily, this is because it is unclear at what spatial scale the lunar thermal-infrared beaming effect is sensitive. Spencer compared his result with surface roughness measurements of the lunar soil made from photographic close-up images taken by the Apollo 11 and 12 astronauts (Lumme, Karttunen & Irvine 1985). He noted that his derived RMS slope of 39º was similar to but greater than the photographically observed roughness of 22 ± 14 degrees RMS slope at 3 mm spatial scales. The results derived in this work are more consistent with this measurement but are still slightly greater. However, it is important to consider the spatial scales that are relevant to the observed fluxes.

The range of spatial scales to which the lunar thermal-infrared beaming effect is sensitive start from the thermal skin depth and end at the spatial resolution of the observations. Considering that the thermal inertia assumed in the best fit model was 50 J m$^{-2}$ K$^{-1}$ s$^{-1/2}$ and that the thermal conductivity measured *in-situ* was ~1 x10$^{-3}$ W m$^{-1}$ K$^{-1}$, gives the lunar thermal skin depth as ~1 cm (using equation 29). The observations were conducted to ~1% lunar radii spatial resolution corresponding to ~18 km. The measured surface roughness therefore has a spatial scale ranging from ~1 cm to ~18 km (a variation of order ~10$^6$). If the thermal skin depth is ~1 cm then the 3 mm spatial scale to which Spencer compared his roughness result is possibly too small. For a relevant comparison, other measurement techniques must be used to determine the degree of surface roughness at ~1 cm scales over at least an 18 km baseline.

Helfenstein & Shepard (1999) utilised images from the Apollo Lunar Surface Closeup Camera (ALSCC) to produce digital topographic relief maps of undisturbed soil of the lunar mare (Apollo 11 and 12) and Fra Mauro regolith (Apollo 14). They measured the 1 cm-scale surface roughness in RMS slope at these regions to be 8.1º ± 2.4º and 12.5º ± 2.0º respectively. This measured degree of surface roughness is much smaller than that implied by the various different thermal models. However, since the close-up images had a footprint of 72 x 82.8 mm the surface roughness analysis was limited to decimetre scales and therefore neglects the roughness statistics at larger scales.

The laser altimeter (LOLA) on the Lunar Reconnaissance Orbiter has recently studied lunar surface roughness at ~1 to 5 m and >50 m scales (Smith et al. 2010). Unfortunately, no data currently exists on lunar surface roughness statistics at ~10 cm to 1 m and ~5 to 50 m scales. If such data did exist then an estimate of lunar surface roughness at 1 cm scales over an 18 km baseline can be obtained by combining the RMS slopes from these studies in quadrature.

Fortunately, lunar surface roughness has also been studied by circular polarised radar observations (Ostro 1993). The derivation of surface roughness from radar data is similar to the thermal infrared beaming method, i.e. it is sensitive to all spatial scales ranging from the observation wavelength to the spot size of the sub-radar point. From lunar radar observations it is estimated that the RMS slope at 1 cm spatial scales is ~33º, which is in precise agreement with that inferred in this work from the lunar thermal-infrared beaming effect.

### 4. APPLICATION TO ASTEROIDS

### 4.1 Investigation Details

Now that the model has been verified by recreating lunar thermal-infrared observations and that the derived surface roughness appears to be consistent with existing lunar radar data, the model is applied to investigate the directional characteristics of asteroid thermal emission. In the following sections the geometrical, wavelength, thermal inertia, and Bond albedo dependencies as a function of observation angle are studied by taking the ratio of rough surface thermal emission to that of a smooth flat surface. This is a huge parameter space to study in detail and so when a specific parameter is studied the other parameters are held constant. To determine the geometrical dependence four illumination geometries are considered: at asteroid midday and midnight ($\theta_{SUN} = 0°$ and $180°$), and near asteroid sunrise and sunset ($\theta_{SUN} = \pm70°$). Finally, the surface power input and output is studied in the presence of surface roughness.

For the investigation a spherical asteroid with a pole orientation perpendicular to its orbital plane and a 6 hour rotation period is assumed to be placed at 1 AU from the Sun. The medium-resolution 90° crater with 50% coverage (i.e. 35° RMS slope) is used to represent unresolved surface roughness for a shape facet placed on the asteroid equator. Table 5 summarises the surface properties used for the investigations.

Figure 6 displays the *ATPM* model results for the various parameters studied.

### 4.2 Input and Output Power

Multiple scattering of sunlight between interfacing facets of a rough surface causes the surface to absorb more sunlight than it normally would if it were smooth and flat. Roughness essentially lowers the effective Bond albedo of the surface, $A_{B\_EFF}$, which for spherical section craters of opening angle $\gamma$ is given by (Müller 2007)

$$A_{B\_EFF} = A_B \frac{1 - \sin^2(\gamma/2)}{1 - A_B \sin^2(\gamma/2)} \,. \tag{43}$$

Figure 7 shows the effective Bond albedo and the corresponding sunlight absorptivity increase (i.e. increase in power input) for a 90° crater as a function of Bond albedo.

Also, re-absorption of emitted thermal radiation between interfacing facets causes the rough surface to heat up and cool down at different rates to those of a smooth flat surface and therefore affects its overall power output. Figure 8 shows the power output for a smooth flat surface and a 90° crater as a function of rotation phase and thermal inertia.

### 4.3 Discussion

Figures 6a and 6b indicate that the thermal-infrared beaming effect is highly wavelength dependent with the shortest wavelengths being beamed the most and the longest wavelengths being beamed the least. The total radiated power integrated over all wavelengths displayed in Figure 6e is also beamed significantly meaning that the overall emitted photon recoil force is generally not perpendicular to the surface. This has implications for predicting the Yarkovsky and YORP effects acting on an asteroid, as all previous models have assumed that the photon recoil force is perpendicular to the surface. The high sensitivity at short wavelengths is dictated by the shift of the steep part of the Planck curve (before the emission peak) towards shorter wavelengths with temperature. The addition of surface roughness causes facets with higher temperatures to become visible to the observer allowing the steep part of the Planck curve to easily shift. It is less sensitive at longer wavelengths because the Planck curve is relatively shallow after the emission peak which shifts less with changes in temperature.

Figure 6c indicates that the beaming effect is thermal inertia dependent with the lowest thermal inertias being beamed the most and the highest thermal inertias being beamed the least. Increasing asymmetry is also observed between the amount of beaming displayed between the morning and afternoon sides of an asteroid with increasing thermal inertia. In the presence of non-zero thermal inertia the morning-side beaming effect is generally higher than the afternoon-side beaming effect.

Figure 6d indicates that there is a slight Bond albedo dependence of the beaming effect which causes the effect to increase with increasing Bond albedo. This is likely to be related to the relative increase in power input with Bond albedo of a rough surface as shown in Figure 7.

All parameter investigations show that the thermal-infrared beaming effect on the sunlit side of an asteroid is highly dependent on the observation and illumination geometry involved. They exhibit the expected result that the beaming effect is greatest when the observation and illumination directions are the same. However, contrary to expectation, the flux enhancement seen in disc-integrated observations of the sunlit side of an asteroid is dominated by limb surfaces rather than the subsolar region. This is clearly shown by the asteroid sunrise and sunset thermal-infrared beaming enhancements being much greater than those at and near asteroid midday. This suggests that for the sunlit side of an asteroid, sunlit surfaces directly facing the observer in situations where they wouldn't be if the surface was a smooth flat one are more important than mutual selfheating between interfacing facets raising their temperatures. Figure 9 demonstrates this effect for a Gaussian random surface during sunrise viewed from different directions. The thermal flux observed is enhanced when viewing hot sunlit surfaces (i.e. Sun behind the observer), and is reduced when viewing cold shadowed surfaces (i.e. Sun in front of the observer).

Jakosky, Finiol & Henderson (1990) also studied the directional thermal emission of Earth-based lava flows exhibiting macroscopic roughness. They found that enhancements in thermal emission were caused by viewing hot sunlit sides of rocks and reductions were caused by viewing cold shadowed sides of rocks. This agrees precisely with the model and adds further evidence that thermal-infrared beaming is caused by macroscopic roughness rather than microscopic roughness.

On the night side of the asteroid the parameter investigations show that the observed thermal emission is enhanced but is not strongly directionally dependent. This suggests that in this case, the mutual selfheating between interfacing facets is more important than viewing them from any particular orientation. Re-absorption of emitted thermal radiation allows the roughness facets to stay hotter for longer because they cool down more slowly. Hot spots on the lunar surface have been observed in thermal-infrared images taken during lunar eclipse which verify this effect (Saari, Shorthill & Deaton 1966). The images clearly show that there are a large number of hot spots corresponding with craters that are warmer than the surrounding terrain.

Related to this, Figure 8 shows that the power output as a function of rotation phase for a 90° crater is enhanced over a smooth flat terrain during the asteroid night, and is consequently reduced during the asteroid day. This is consistent with the enhanced thermal emissions observed on the night side of the asteroid. The day-side power outputs have to be reduced to maintain energy conservation and this is seen as the reduction in thermal emission at high phase angles. By comparing the power output curves for different thermal inertias it appears that surface roughness increases the effective thermal inertia of the surface i.e. it acts like an additional energy storage device. This has implications for predicting the magnitude of the Yarkovsky effect on an asteroid since it is highly dependent on thermal inertia. As mentioned before, all previous Yarkovsky models have neglected surface roughness and its thermal-infrared beaming effect.

## 5. SUMMARY AND CONCLUSIONS

The implementation of a new thermophysical model called the Advanced Thermophysical Model (*ATPM*) is described. It is an improvement over previous thermophysical models as it includes partial shadowing and visibility techniques to allow more accurate calculation of thermal emission at high observation angles, and better viewfactor calculations to allow any type of surface roughness model to be used. It also includes global-selfheating effects which previous models have neglected.

The rough surface thermal model accurately reproduces the lunar thermal-infrared beaming effect at a surface roughness of ~32º RMS slope by assuming surface thermal properties that have been measured *in-situ*. The derived surface roughness is almost independent of how it is represented in a topography model. However, the topography model must have sufficient surface roughness in order to ensure its maximum thermal-infrared beaming effect is greater than or equal to that observed. The derived surface roughness is an accumulation of roughness at all spatial scales ranging from the thermal skin depth to the spatial resolution of the observations, and is consistent with lunar surface roughness measured by radar.

By considering the huge range of potential asteroid surface properties, the rough-surface model implies a thermal-infrared beaming effect that cannot be described by a simple parameter or function. The beaming effect was found to be highly dependent on the observation and illumination geometry, and also the surface thermal properties. Contrary to expectation, the flux enhancement seen in disc-integrated observations is dominated by limb surface enhancements rather than enhancements from the subsolar region. For accurate determination of asteroid surface thermal properties, surface roughness must be explicitly modelled and preferably aided with thermal measurements conducted at a number of different wavelengths and made at a number of different phase angles.

It was also found that thermal-infrared beaming is predominantly caused by macroscopic rather than microscopic roughness. On the asteroid day side hot sunlit surfaces facing the observer are most important, whilst on the asteroid night side it is the mutual selfheating of interfacing surface elements. The inclusion of microscopic beaming has minimal effect in the predicted directional thermal emission and for simplicity purposes can be neglected from thermophysical models.

Finally, surface roughness and its associated thermal-infrared beaming effect moves the overall emission angle of thermal flux away from the surface normal, and alters the effective Bond albedo and thermal inertia of the surface. This has implications for predicting the Yarkovsky and YORP effects acting on asteroids which are highly dependent on those properties. Since previous Yarkovsky and YORP models have neglected these effects, their impact on the predictions has been studied in more detail in an accompanying paper (Rozitis & Green 2010).

**Acknowledgements**

We are grateful to the reviewer Dr. A. W. Harris for several suggested refinements to the manuscript. The work of BR is supported by the UK Science and Technology Facilities Council (STFC).

# APPENDIX A: Coordinate Systems Geometry

Figures A1 and A2 depict the five coordinate systems mentioned in section 2.5. The co-rotating planetcentric equatorial system ($x_0$, $y_0$, $z_0$) defines the global shape of the planetary body and can be transformed into the planetcentric equatorial system ($x_{equ}$, $y_{equ}$, $z_{equ}$) by a rotational transformation. The $x_0$ axis is aligned with the planetary prime meridian. The two systems are separated by an angle $\omega t$ where $\omega$ is the planetary angular rotation rate and $t$ is the time since an initial epoch when rotations are considered to have begun. The transformation is given by

$$
\begin{aligned}
x_{equ} &= x_0 \cos \omega t - y_0 \sin \omega t \\
y_{equ} &= x_0 \sin \omega t + y_0 \cos \omega t \\
z_{equ} &= z_0
\end{aligned}
\tag{A1}
$$

The planetcentric ecliptic coordinate system ($x_{ecli}$, $y_{ecli}$, $z_{ecli}$) takes into account the rotation pole orientation specified by the polar coordinates $\lambda_P$ and $\beta_P$ which are the planetcentric ecliptic longitude and latitude respectively. It is specified such that the $x_{ecli}$ axis has a component in the direction of the first point of Aries allowing the planetary prime meridian to align also with this point at time zero. The planetcentric equatorial and planetcentric ecliptic are related by the following sets of transformations

$$
\begin{bmatrix} x_{ecli} \\ y_{ecli} \\ z_{ecli} \end{bmatrix} = \begin{bmatrix} u_x & v_x & w_x \\ u_y & v_y & w_y \\ u_z & v_z & w_z \end{bmatrix} \begin{bmatrix} x_{equ} \\ y_{equ} \\ z_{equ} \end{bmatrix}
\tag{A2}
$$

$$
\begin{bmatrix} x_{equ} \\ y_{equ} \\ z_{equ} \end{bmatrix} = \begin{bmatrix} u_x & u_y & u_z \\ v_x & v_y & v_z \\ w_x & w_y & w_z \end{bmatrix} \begin{bmatrix} x_{ecli} \\ y_{ecli} \\ z_{ecli} \end{bmatrix}
\tag{A3}
$$

where $u_i$, $v_i$, and $w_i$ are components of the unit vectors representing the planetcentric equatorial system when inside the planetcentric ecliptic frame of reference. The $u_i$, $v_i$, and $w_i$ components are given by

$$
\begin{aligned}
u_x &= \frac{\sin \beta_P}{\sin \alpha} \\
u_y &= 0 \\
u_z &= \frac{-\cos \beta_P \cos \lambda_P}{\sin \alpha}
\end{aligned}
\tag{A4}
$$

$$
\begin{aligned}
v_x &= \frac{-\cos^2 \beta_P \sin \lambda_P \cos \lambda_P}{\sin \alpha} \\
v_y &= \frac{\sin^2 \beta_P + \cos^2 \beta_P \cos^2 \lambda_P}{\sin \alpha} \\
v_z &= \frac{-\sin \beta_P \cos \beta_P \sin \lambda_P}{\sin \alpha}
\end{aligned}
\tag{A5}
$$

$$
\begin{aligned}
w_x &= \cos \beta_P \cos \lambda_P \\
w_y &= \cos \beta_P \sin \lambda_P \\
w_z &= \sin \beta_P
\end{aligned}
\tag{A6}
$$

where

$$\alpha = \cos^{-1}\left(\cos\beta_P \sin|\lambda_P|\right) . \tag{A7}$$

The planetcentric ecliptic system can be converted to the heliocentric ecliptic system ($x_H$, $y_H$, $z_H$) by taking into account the position of the planetary body with respect to the Sun, ecliptic plane, and the first point of Aries. If a planetary body has heliocentric coordinates $r_H$, $\lambda_H$, and $\beta_H$ then

$$\begin{aligned} x_H &= r_H \cos\beta_H \cos\lambda_H + x_{ecli} \\ y_H &= r_H \cos\beta_H \sin\lambda_H + y_{ecli} \\ z_H &= r_H \sin\beta_H + z_{ecli} \end{aligned} \tag{A8}$$

Finally, since the model is intended to utilise any type of surface topography it is convenient to define an additional coordinate system for the unresolved surface roughness. In this coordinate system ($x_S$, $y_S$, $z_S$), new surface topography shape models can be generated, and thermal model calculations can be performed by transforming the appropriate global shape model geometry into this system. The $x_S$ and $y_S$ axes define a plane that would lie parallel to the plane of a shape facet with the $x_S$ axis lying parallel with the shape facet's vector $\bm{p_1} - \bm{p_0}$. The $z_S$ axis is therefore perpendicular to this plane and lies parallel with the shape facet normal. For determining angles of interest (e.g. illumination and observation angles) between the roughness facet normals and a vector specified in one of the external coordinate systems defined above, the vector of interest must first be transformed into the surface-roughness coordinate system. These two coordinate systems are related by the following transformations

$$\begin{bmatrix} x_S \\ y_S \\ z_S \end{bmatrix} = \begin{bmatrix} u_x & u_y & u_z \\ v_x & v_y & v_z \\ w_x & w_y & w_z \end{bmatrix} \begin{bmatrix} x \\ y \\ z \end{bmatrix} \tag{A9}$$

$$\begin{bmatrix} x \\ y \\ z \end{bmatrix} = \begin{bmatrix} u_x & v_x & w_x \\ u_y & v_y & w_y \\ u_z & v_z & w_z \end{bmatrix} \begin{bmatrix} x_S \\ y_S \\ z_S \end{bmatrix} \tag{A10}$$

where $x$, $y$, and $z$ are the components of the vector of interest in the external coordinate system, and the $u_i$, $v_i$, and $w_i$ components in this case are given by

$$\begin{aligned} u_x &= \frac{p_{1,x} - p_{0,x}}{|\bm{p_1} - \bm{p_0}|} \\ u_y &= \frac{p_{1,y} - p_{0,y}}{|\bm{p_1} - \bm{p_0}|} \\ u_z &= \frac{p_{1,z} - p_{0,z}}{|\bm{p_1} - \bm{p_0}|} \end{aligned} \tag{A11}$$

$$\begin{aligned} v_x &= \frac{n_y(p_{1,z} - p_{0,z}) - n_z(p_{1,y} - p_{0,y})}{|\bm{p_1} - \bm{p_0}|} \\ v_y &= \frac{n_z(p_{1,x} - p_{0,x}) - n_x(p_{1,z} - p_{0,z})}{|\bm{p_1} - \bm{p_0}|} \\ v_z &= \frac{n_x(p_{1,y} - p_{0,y}) - n_y(p_{1,x} - p_{0,x})}{|\bm{p_1} - \bm{p_0}|} \end{aligned} \tag{A12}$$

$$w_x = n_x$$
$$w_y = n_y \quad (A13)$$
$$w_z = n_z$$

where $p_{0,i}$ and $p_{1,i}$ are the components of position vectors $\boldsymbol{p_0}$ and $\boldsymbol{p_1}$, and $n_i$ are the components of the unit normal vector $\boldsymbol{n}$ of the shape facet in the external coordinate system. Depending on how the surface topography model is generated it could have different spatial units to the global shape model, and therefore a different projected area in the plane of the shape facet for which it is representing unresolved surface roughness. An area conversion factor is required to be applied to any calculation that involves area (e.g. determining the observed surface thermal emission). The area conversion factor $ACF$ is given by

$$ACF = a \bigg/ \sum_{i=1}^{M} a_i n_{z,i} \quad (A14)$$

where $a$ is the surface area of the shape facet, and $a_i$ is the area and $n_{z,i}$ is the $z_S$ axis component of the unit normal of roughness facet $i$ (for $i = 1$ to $M$ roughness facets).

**Tables**

*Table 1: Shape properties of the various rough surfaces used in this work.*

| Roughness Variant | Number of Vertices | Number of Facets | RMS Slope / ° | Maximum Slope / ° | Mean Total Viewfactor |
|---|---|---|---|---|---|
| Smooth Flat Surface | 3 | 1 | 0.0 | 0.0 | 0.000 |
| 30° Crater | 613 | 1188 | 20.9 | 29.6 | 0.067 |
| 45° Crater | 721 | 1404 | 30.7 | 44.6 | 0.147 |
| 60° Crater | 829 | 1620 | 39.3 | 59.6 | 0.251 |
| High-Res. 90° Crater | 1045 | 2052 | 49.1 | 89.5 | 0.501 |
| Med-Res. 90° Crater | 325 | 612 | 49.2 | 85.0 | 0.500 |
| Low-Res. 90° Crater | 73 | 132 | 50.0 | 82.8 | 0.510 |
| High-Res. Gaussian | 1089 | 2048 | 49.1 | 78.7 | 0.348 |
| Low-Res. Gaussian | 121 | 200 | 35.9 | 64.8 | 0.173 |

*Table 2: Lunar surface model parameters.*

| Parameter | Value |
|---|---|
| Heliocentric Position | $r_H = 1$ AU, $\lambda_H = 0°$, $\beta_H = 0°$ |
| Solar Flux, $F_{SUN}$ | 1360 W m$^{-2}$ |
| Pole Orientation | $\lambda_P = 0°$, $\beta_P = 90°$ |
| Rotation Period, $P$ | 2551440.0 s |
| Bond Albedo, $A_B$ | 0.1 |
| Emissivity, $\varepsilon$ | 0.9 |
| Thermal Albedo, $A_{TH}$ | 0.1 |
| Thermal Inertia, $\Gamma$ | 50 J m$^{-2}$ K$^{-1}$ s$^{-1/2}$ |
| Convergence Goal, $T_{ACC}$ | 0.05 K |

*Table 3: Lunar model rough surface fitting results.*

| Roughness Variant | Varying Sun Angle | | Varying Observer Angle | |
|---|---|---|---|---|
| | $\chi^2$ | RMS Slope / ° | $\chi^2$ | RMS Slope / ° |
| Smooth Flat Surface | 1.000 | 0.0 | 1.000 | 0.0 |
| 30º Crater | 0.453 | >20.9 | 0.24 | >20.9 |
| 45º Crater | 0.253 | 27.9 ± 2.7 | 0.13 | 27.3 ± 2.5 |
| 60º Crater | 0.190 | 30.2 ± 2.5 | 0.10 | 30.5 ± 1.8 |
| High-Res. 90º Crater | 0.201 | 32.2 ± 2.6 | 0.098 | 34.0 ± 1.9 |
| Med-Res. 90º Crater | 0.183 | 32.7 ± 2.4 | 0.097 | 34.5 ± 1.9 |
| Low-Res. 90º Crater | 0.180 | 33.1 ± 2.6 | 0.098 | 35.3 ± 2.0 |
| High-Res. Gaussian | 0.222 | 37.1 ± 3.6 | 0.128 | 39.6 ± 2.7 |
| Low-Res. Gaussian | 0.405 | 29.4 ± 4.7 | 0.185 | 32.9 ± 2.9 |
| | **Average** | 31.5 ± 1.5 | | 33.0 ± 1.1 |

*Table 4: Lunar surface roughness derived by various thermal models.*

| Model | Derived RMS Slope / ° |
|---|---|
| Buhl, Welch & Rea (1968) | 35 |
| Sexl et al. (1971) | 30 |
| Winter & Krupp (1971) | 34 |
| Spencer (1990) | 39 |
| Shkuratov et al. (2000) | 30 |
| This work (varying sun angle) | 31.5 ± 1.5 |
| This work (varying observer angle) | 33.0 ± 1.1 |

*Table 5: Assumed surface properties for parameter investigation of ATPM applied to a test asteroid.*

| Investigation | Wavelength / μm | Thermal Inertia / J m$^{-2}$ K$^{-1}$ s$^{-1/2}$ | Bond Albedo |
|---|---|---|---|
| Wavelength | 2.5, 5.0, 10, All | 200 | 0.1 |
| Thermal Inertia | 10 | 0, 200, 750, 1500 | 0.1 |
| Bond Albedo | 10 | 200 | 0.1, 0.3, 0.5 |

**Figure Captions**

Figure 1: Directional illumination and observation geometry.

Figure 2: Schematic of the Advanced Thermophysical Model (*ATPM*) where the terms $F_{SUN}$, $F_{SCAT}$, $F_{RAD}$, $k(dT/dx)$, and $\varepsilon\sigma T^4$ are the direct sunlight, multiple scattered sunlight, reabsorbed thermal radiation, conducted heat, and thermal radiation lost to space respectively.

Figure 3: Wireframe renderings of various rough surfaces. (1st row) 30º and 45º craters. (2nd row) 60º crater and 90º high resolution crater. (3rd row) 90º medium resolution and low resolution craters. (4th row) High resolution and low resolution Gaussian random height surfaces.

Figure 4: Best model fit (lines) for the medium-resolution 90º crater to observed lunar surface brightness temperatures (circles and triangles). (a) Observation angles of -30º (triangles and dashed line) and -53º (circles and solid line). (b) Observation angles of +30º (triangles and dashed line) and +53º (circles and solid line). (c) Observation angle of 0º.

Figure 5: Best model fit for the medium-resolution 90º crater to observed lunar direction factors. The triangles and dashed lines correspond to lunar morning observations and model fits respectively, and the circles and solid lines correspond to the lunar afternoon.

Figure 6: Parameter dependence of directionally resolved thermal-infrared flux ratios predicted by *ATPM* for a rough asteroid surface. (a) Wavelength dependence at asteroid midday ($\theta_{SUN} = 0°$) and midnight ($\theta_{SUN} = 180°$). The solid, dashed, and dotted lines correspond to observation wavelengths of 2.5, 5.0, and 10 μm respectively. (b) Wavelength dependence near asteroid sunrise ($\theta_{SUN} = -70°$) and sunset ($\theta_{SUN} = +70°$). The solid, dashed, and dotted lines correspond to observation wavelengths of 2.5, 5.0, and 10 μm respectively. (c) Thermal inertia depedence near asteroid sunrise and sunset. The solid, dashed, dotted, and dash-dotted lines correspond to surface thermal inertias of 0, 200, 750, and 1500 J m$^{-2}$ K$^{-1}$ s$^{-1/2}$ respectively. (d) Bond albedo dependence near asteroid sunrise and sunset. The solid, dashed, and dotted lines correspond to Bond albedos of 0.1, 0.3, and 0.5 respectively. (e) Directionally resolved dependence of total radiated power integrated over all wavelengths as a function of the different sun illumination angles given in the top right corner.

Figure 7: Effective Bond albedo and absorptivity increase for a 90º crater as a function of Bond albedo. Effective Bond albedo is given by the primary y-axis and the absorptivity increase is given by the secondary y-axis with the line representing both.

Figure 8: Power output as a function of rotation phase and thermal inertia (represented by the different line styles as indicated in the top right corner). Thick lines represent a smooth flat surface and the thin lines represent the 90º crater.

Figure 9: Sunrise surface temperatures for a Gaussian random height surface viewed from different directions. The black line gives the Sun direction and the colour bar scale indicates the surface temperatures derived in the model.

Figure A1: Model coordinate systems.

Figure A2: Surface-roughness coordinate system.

**Figures**

*Figure 1*

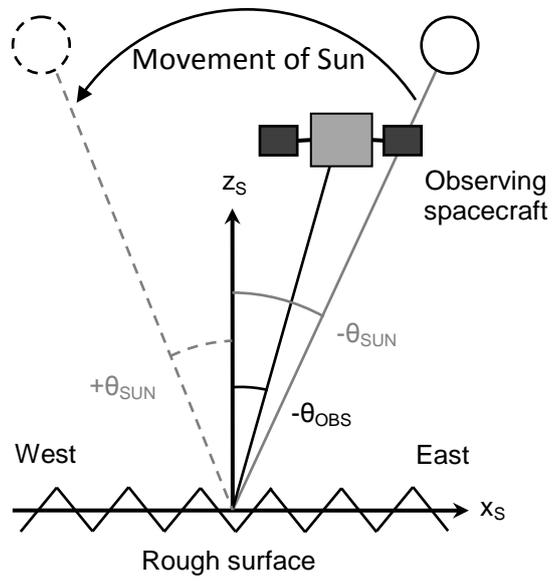

*Figure 2*

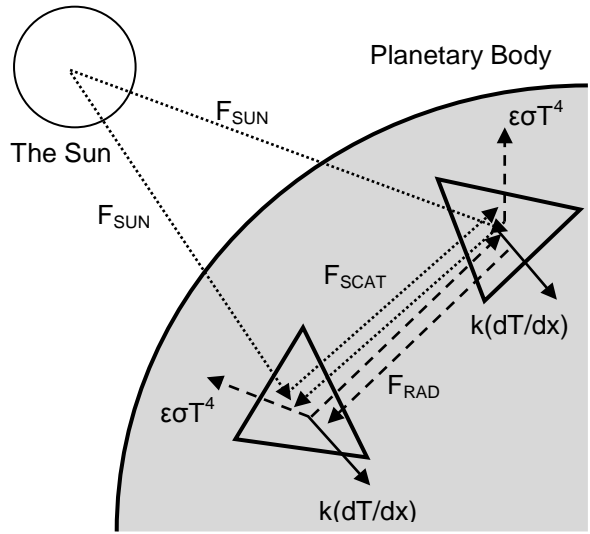

*Figure 3*

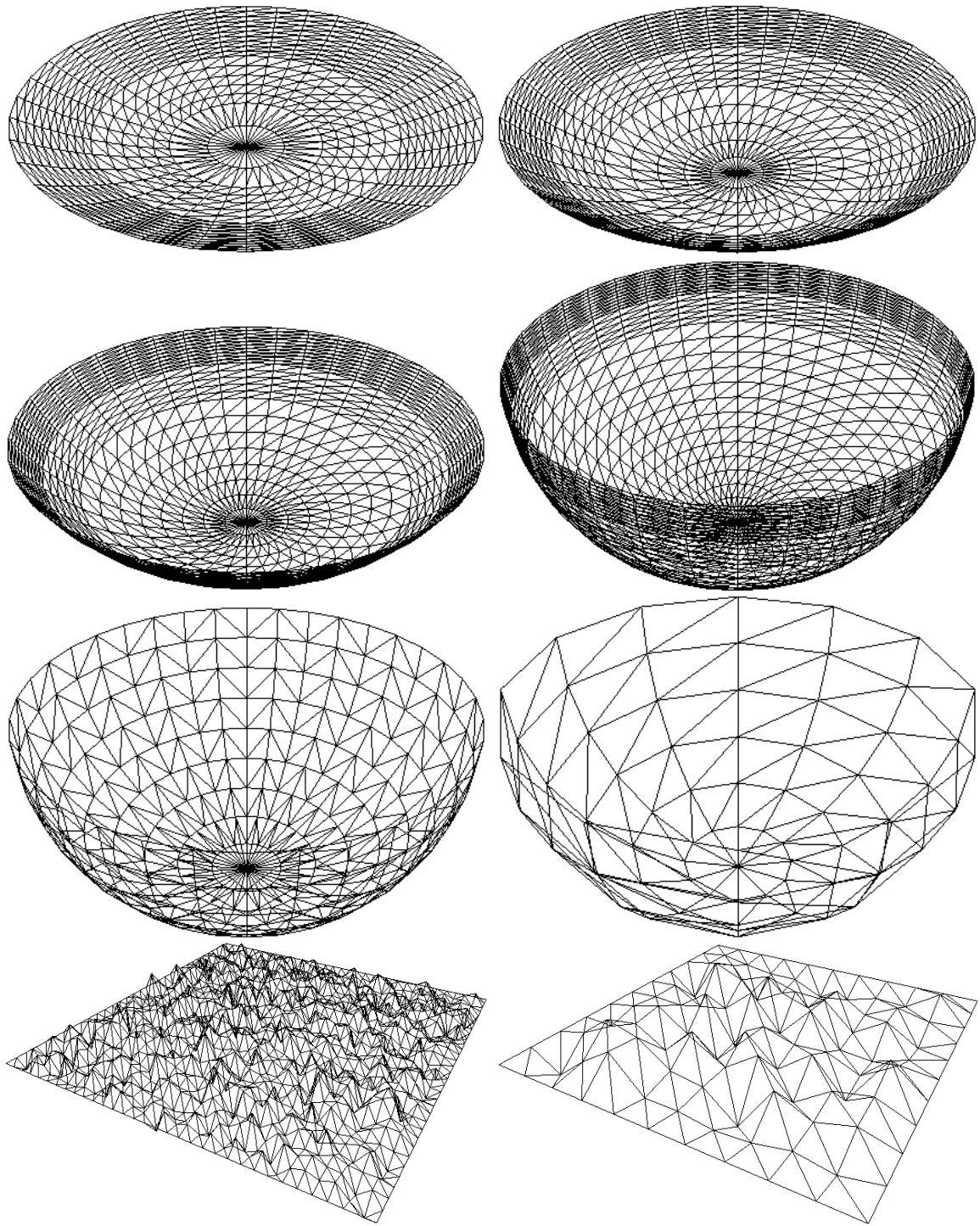

*Figure 4*

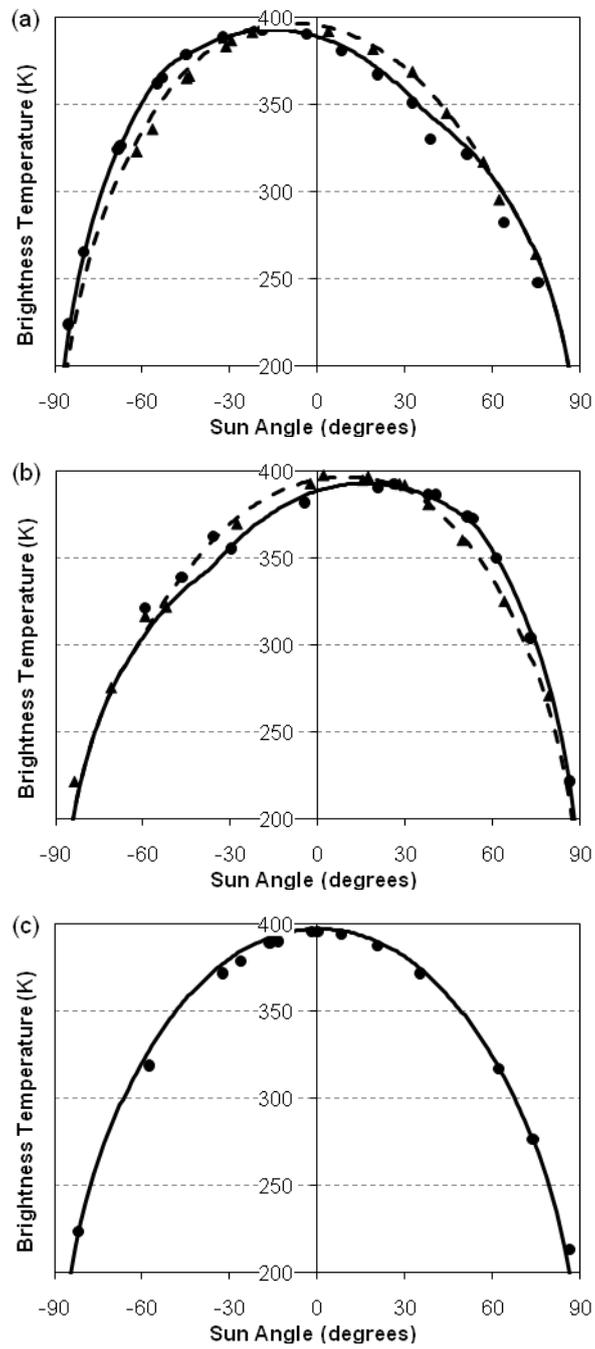

*Figure 5*

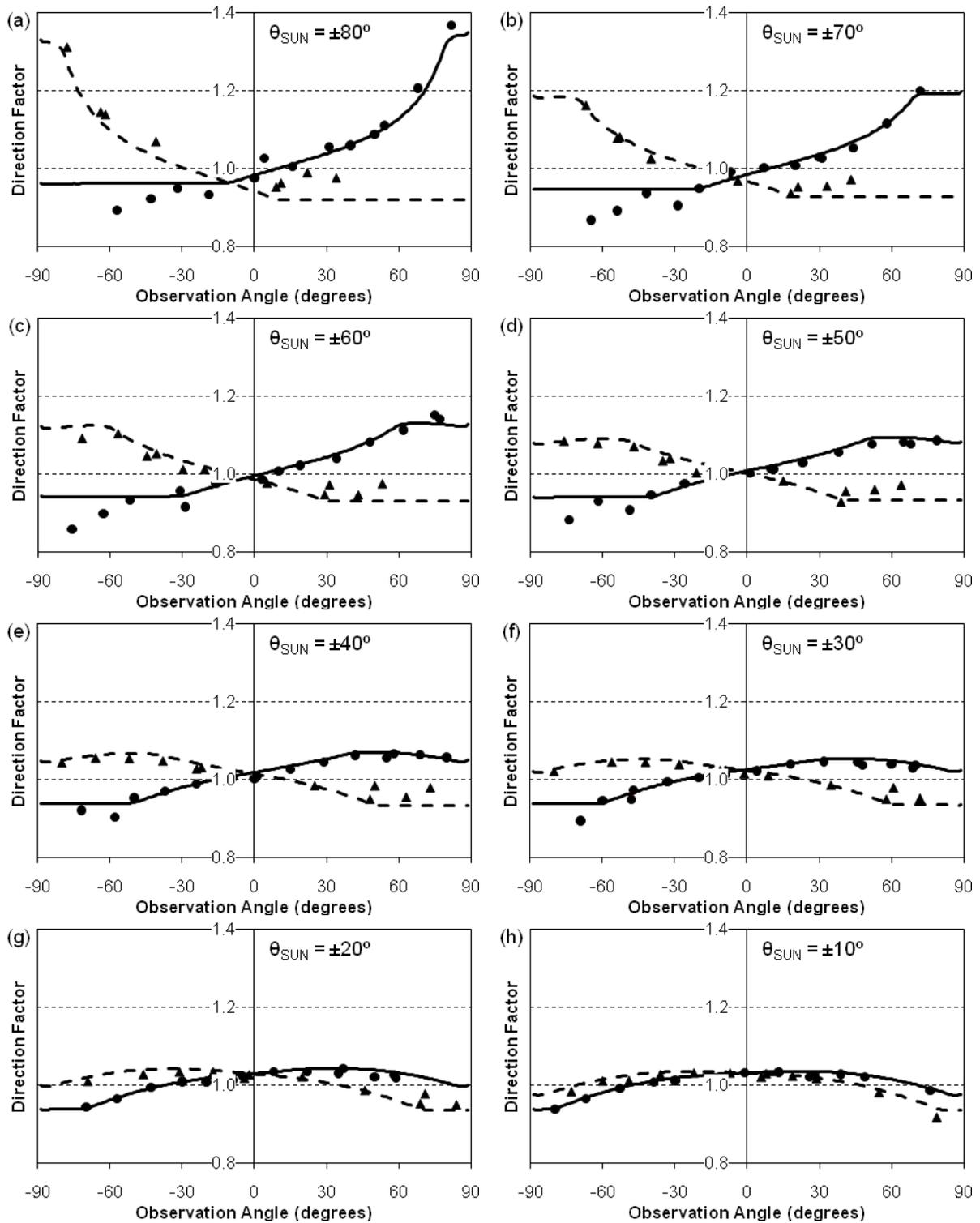

*Figure 6*

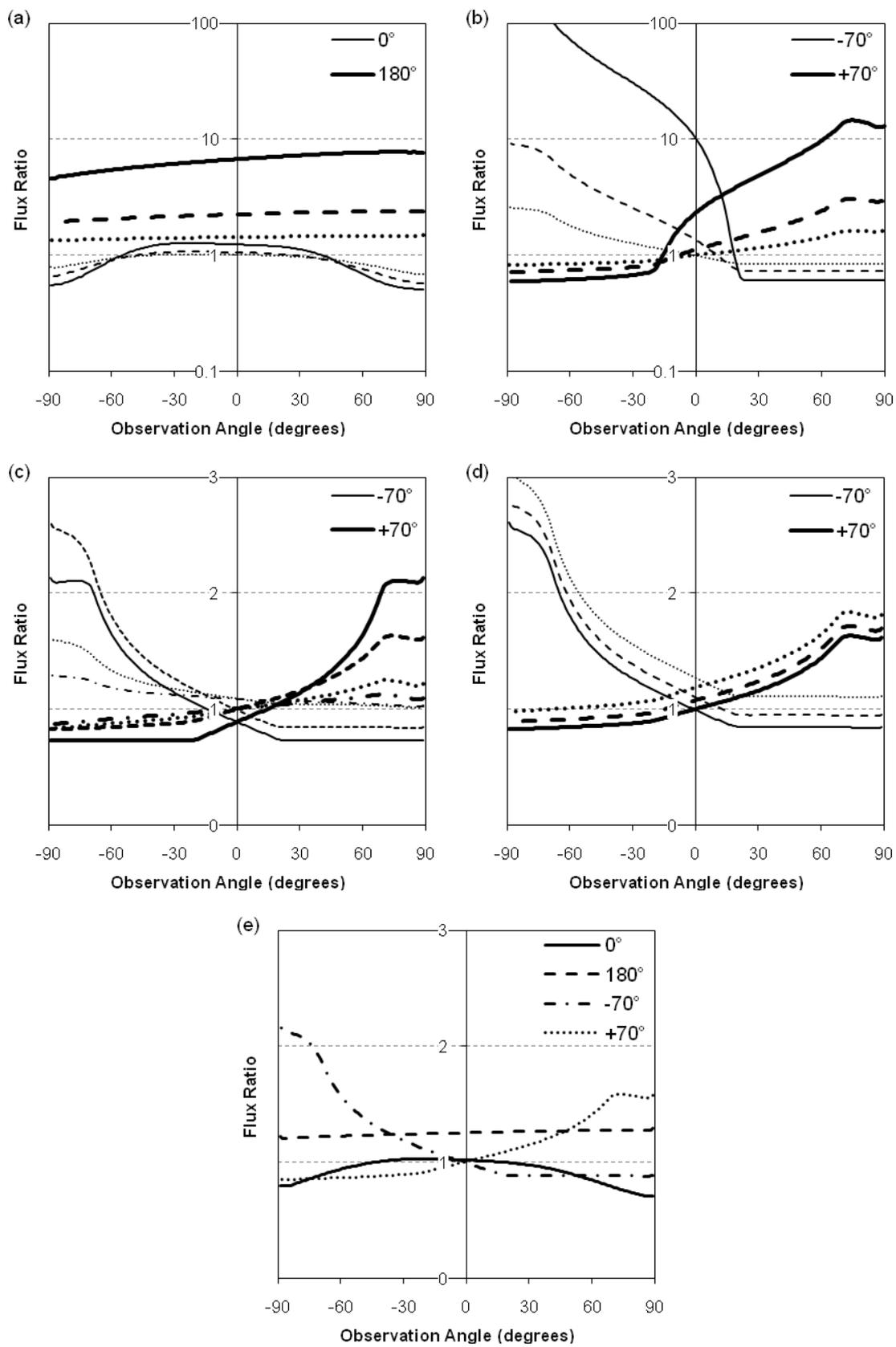

*Figure 7*

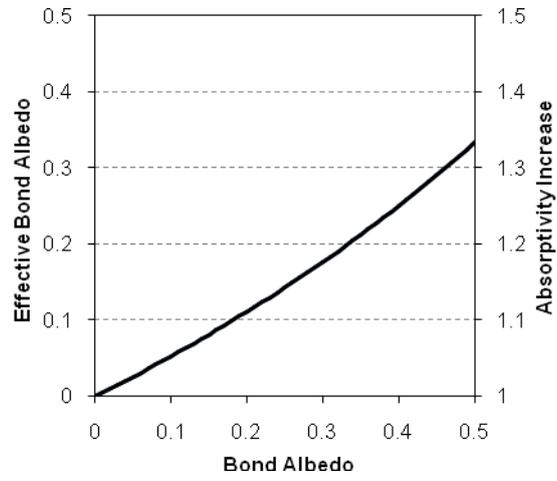

*Figure 8*

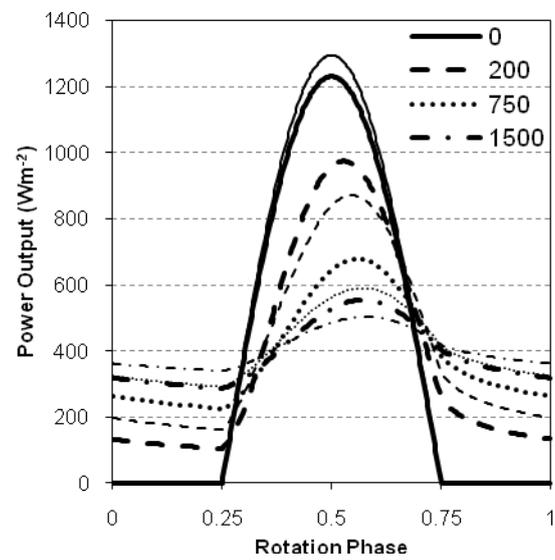

*Figure 9*

(Colour version for reproduction on the web only)

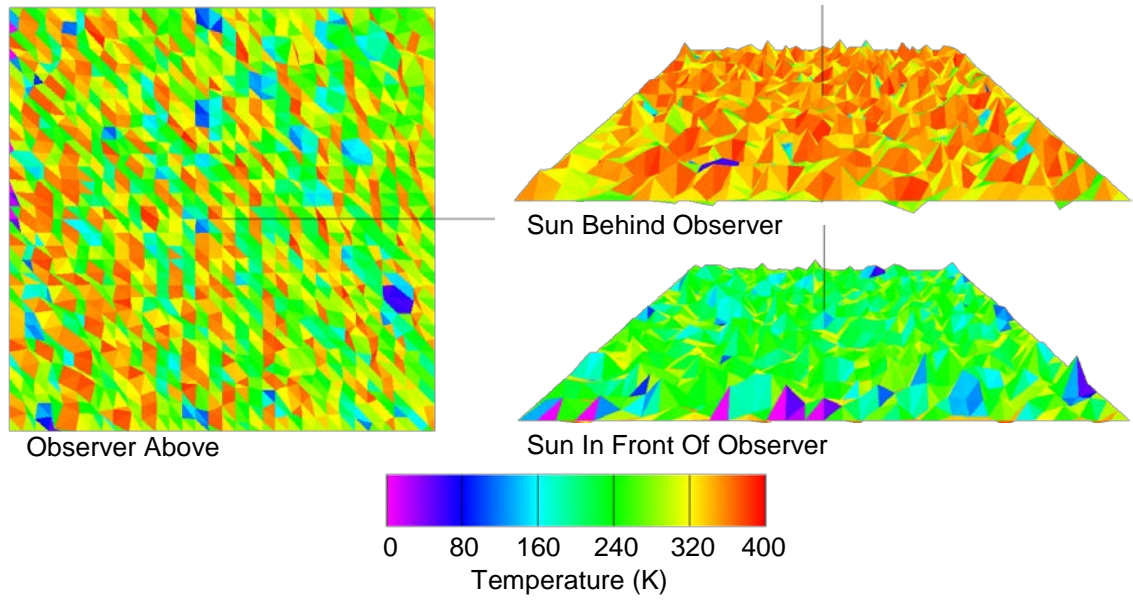

(Greyscale version for print publication)

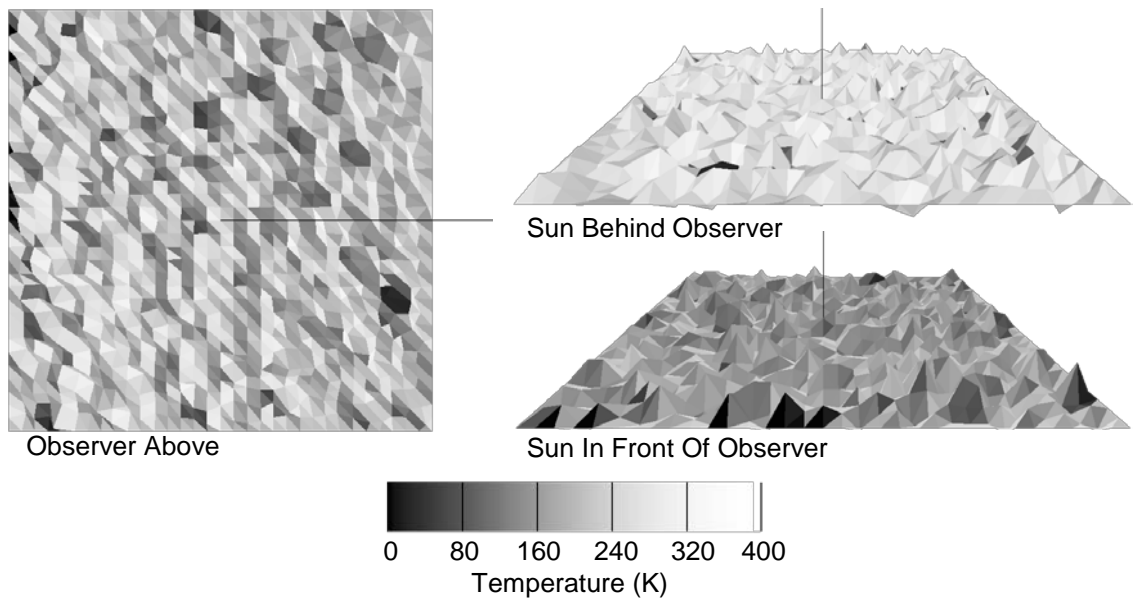

*Figure A1*

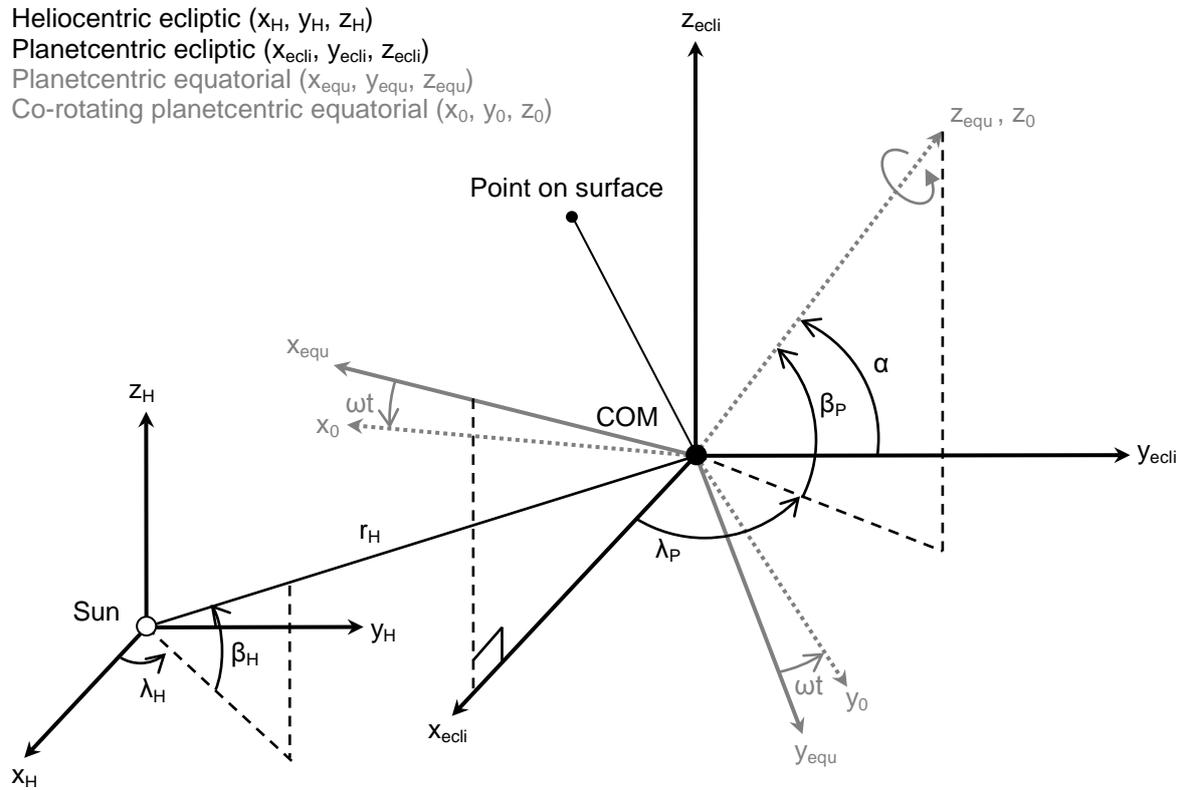

*Figure A2*

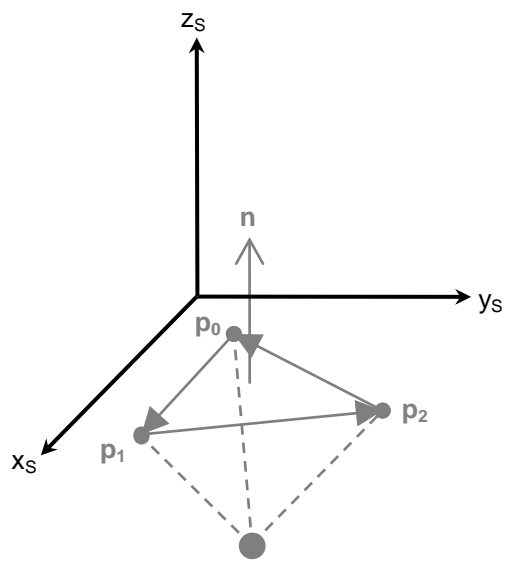